# One-loop renormalization of the QCD Schrödinger functional

Stefan Sint

Max-Planck-Institut für Physik
– Werner-Heisenberg-Institut –
Föhringer Ring 6
D-80805 München, Germany

## Abstract

In a previous publication, we have constructed the Schrödinger functional in Wilson's lattice QCD. It was found that the naive continuum limit leads to a well-defined classical continuum theory. Starting from the latter, a formal continuum definition of the Schrödinger functional is given and its saddle point expansion is carried out to one-loop order of perturbation theory. Dimensional regularization and heat kernel techniques are used to determine the one-loop divergences. These are partly canceled by the usual renormalizations of the quark mass and the coupling constant in QCD. An additional divergence can be absorbed in a multiplicative renormalization of the quark boundary fields. The corresponding boundary counterterm is a local polynomial in the fields, so that we confirm a general expectation of Symanzik.

April 1995

# 1. Introduction

The Schrödinger functional in a quantum field theory is defined as the transition amplitude between field configurations at (euclidean) times $x_0 = 0$ and $x_0 = T$. It can be written as a path integral, where only those euclidean fields are integrated over which assume the initial and final field configurations as boundary values. In other words, the Schrödinger functional can be regarded as a euclidean quantum field theory, defined on a space-time manifold with boundaries and (inhomogeneous) Dirichlet boundary conditions for the quantum fields.

Perturbative renormalizability of a quantum field theory is usually established using power counting in momentum space. In the case of the Schrödinger functional, this is not possible because translation invariance in the time direction is violated. Therefore, one has to worry about the universality of the Schrödinger functional even in perturbation theory. At this point it is useful to recall Symanzik's work on the Schrödinger representation in quantum field theory [1,2]. He considered the case of massless scalar $\phi^4$ field theory and found that the Schrödinger functional could be renormalized by adding two new counterterms to the action. These are proportional to the local composite fields $\phi\partial_0\phi$ and $\phi^2$, integrated over the hyper-planes at $x_0 = 0$ and $x_0 = T$.

Following Symanzik, it is plausible that this result is generic for a renormalizable quantum field theory. We thus expect that the Schrödinger functional in a renormalizable quantum field theory is again renormalizable, after inclusion of a finite number of additional boundary counterterms. These are local polynomials in the fields and derivatives thereof, integrated over the boundary. Furthermore, they must respect the symmetries of the theory and have canonical dimension less than or equal to 3. In the SU(N) Yang-Mills theory, this expectation was confirmed to one-loop order of perturbation theory [3]. It turns out that no additional counterterm is needed in this case, basically because there exists no local gauge invariant composite field of dimension 3 or less, if invariance under parity is assumed.

It is the aim of this paper to show how renormalization works out for the Schrödinger functional in QCD. In a previous publication the Schrödinger functional has been defined in Wilson's lattice QCD [4]. It was shown that the naive continuum limit of the lattice action leads to a sensible classical continuum theory. Taking the classical theory as starting point, we will use dimensional regularization to analyze the QCD Schrödinger functional to one-loop order of perturbation theory.

The paper is organized as follows. Sect.2 reviews some aspects of the classical continuum theory. In sect.3, a formal continuum approach to the Schrödinger functional prepares the ground for the saddle point expansion, which is carried out in sects.4 and 5. We then discuss the renormalization of the Schrödinger functional and draw our conclusions. Finally, three appendices have been included. Appendix A summarizes some of the notational conventions. In appendix B, the heat kernels associated with the fluctuation operators are defined and some explicit expressions are given for the case of vanishing background field. Appendix C supplies an asymptotic expansion which is needed to extract the divergent part of the quark self-energy.

## 2. Classical Theory

In ref.[4], a classical continuum action was derived from the lattice Schrödinger functional and some aspects of the corresponding classical theory have already been discussed there. It is worthwhile to have a further look at the classical theory since it provides the starting point for the formulation of the Schrödinger functional in dimensional regularization. In particular, we construct explicit expressions for the classical solutions of the Dirac equations, which will later play the rôle of test functions.

*The classical action*

The gauge part of the classical continuum action is given by

$$S_g[A] = -\frac{1}{2g_0^2} \int_0^T dx_0 \int_0^L d^3\mathbf{x}\, \text{tr}\,\{F_{\mu\nu}F_{\mu\nu}\}, \tag{2.1}$$

where $F_{\mu\nu}$ is the field strength tensor associated to the SU($N$) gauge field $A_\mu$,

$$F_{\mu\nu} = \partial_\mu A_\nu - \partial_\nu A_\mu + [A_\mu, A_\nu], \tag{2.2}$$

and the spatial vector components of the gauge field are required to satisfy the boundary conditions

$$A_k(x)|_{x_0=0} = C_k(\mathbf{x}), \qquad A_k(x)|_{x_0=T} = C'_k(\mathbf{x}). \quad k = 1,2,3. \tag{2.3}$$

It will be assumed that there are $n_f$ degenerate quark flavors of mass $m$, with action

$$S_f[A,\bar\psi,\psi] = \int_0^T dx_0 \int_0^L d^3\mathbf{x}\,\bar\psi(x)(\slashed{D}+m)\psi(x) \\ - \int_0^L d^3\mathbf{x}\,\left[\bar\psi(x)P_-\psi(x)\right]_{x_0=0} - \int_0^L d^3\mathbf{x}\,\left[\bar\psi(x)P_+\psi(x)\right]_{x_0=T}. \tag{2.4}$$

The covariant derivative is $D_\mu = \partial_\mu + A_\mu$ and the projectors $P_\pm = \frac{1}{2}(1\pm\gamma_0)$ are used to project on the Dirichlet-components of the quark fields. More precisely, the boundary conditions in the euclidean time direction are

$$\begin{aligned} P_+\psi(x)|_{x_0=0} &= \rho_+(\mathbf{x}), & P_-\psi(x)|_{x_0=T} &= \rho'_-(\mathbf{x}), \\ \bar\psi(x)P_-|_{x_0=0} &= \bar\rho_-(\mathbf{x}), & \bar\psi(x)P_+|_{x_0=T} &= \bar\rho'_+(\mathbf{x}), \end{aligned} \tag{2.5}$$

whereas, in the spatial directions, periodic boundary conditions with period $L$ are assumed for all fields.

As pointed out in ref.[4], the form of the fermionic action follows unambiguously from the boundary conditions (2.5), if one assumes parity invariance of the action † and the existence of smooth classical solutions to the equations of motion.

*Classical solutions*

The classical solutions $\psi_{cl}$ and $\bar\psi_{cl}$ are determined through the boundary quark fields. In order to construct them explicitly, one needs the classical quark propagator, defined through
$$(\not{D}+m)S(x,x') = \delta(x-x'), \tag{2.6}$$
and the boundary conditions
$$\begin{aligned}P_+ S(x,x')|_{x_0=0} &= 0, & P_- S(x,x')|_{x_0=T} &= 0, \\ S(x,x')P_-\big|_{x'_0=0} &= 0, & S(x,x')P_+\big|_{x'_0=T} &= 0.\end{aligned} \tag{2.7}$$

In ref.[4], it has been shown that this propagator is well defined if the gauge potential vanishes. It will be assumed here that this remains true in presence of a sufficiently well-behaved gauge potential. For the classical solutions one then obtains
$$\begin{aligned}\psi_{cl}(x) &= \int \mathrm{d}^3\mathbf{x}' \left[ S(x;0,\mathbf{x}')\rho_+(\mathbf{x}') + S(x;T,\mathbf{x}')\rho'_-(\mathbf{x}') \right], \\ \bar\psi_{cl}(x) &= \int \mathrm{d}^3\mathbf{x}' \left[ \bar\rho_-(\mathbf{x}')S(0,\mathbf{x}';x) + \bar\rho'_+(\mathbf{x}')S(T,\mathbf{x}';x) \right].\end{aligned} \tag{2.8}$$

After partial integration in eq.(2.4), the action of the classical fields assumes the more symmetric form,
$$S_f[A,\bar\psi_{cl},\psi_{cl}] = -\tfrac{1}{2}\int_0^L \mathrm{d}^3\mathbf{x}\left[\bar\psi_{cl}\psi_{cl}\right]_{x_0=0} - \tfrac{1}{2}\int_0^L \mathrm{d}^3\mathbf{x}\left[\bar\psi_{cl}\psi_{cl}\right]_{x_0=T}, \tag{2.9}$$

and, using eq.(2.8), a functional of the boundary quark fields is obtained,
$$\begin{aligned}S_f[A,\bar\psi_{cl},\psi_{cl}] = -\int \mathrm{d}^3\mathbf{x}\,\mathrm{d}^3\mathbf{x}' \Big[ &\bar\rho_-(\mathbf{x})S(0,\mathbf{x};T,\mathbf{x}')\rho'_-(\mathbf{x}') \\ &+ \bar\rho_-(\mathbf{x})S(0,\mathbf{x};0,\mathbf{x}')\rho_+(\mathbf{x}') \\ &+ \bar\rho'_+(\mathbf{x})S(T,\mathbf{x};T,\mathbf{x}')\rho'_-(\mathbf{x}') \\ &+ \bar\rho'_+(\mathbf{x})S(T,\mathbf{x};0,\mathbf{x}')\rho_+(\mathbf{x}') \Big].\end{aligned} \tag{2.10}$$

---

† Parity invariance of the QCD action is assumed throughout the paper. In particular the vacuum angle $\theta$ [5,6] is taken to vanish.

## 3. Formal continuum approach

In the continuum, the Schrödinger functional can be represented as a formal euclidean path integral [3],

$$\mathcal{Z}[\bar{\rho}'_+, \rho'_-, C'; \bar{\rho}_-, \rho_+, C] = \int D[\Lambda] \int D[\psi]D[\bar{\psi}]D[A]\, e^{-S[A, \bar{\psi}, \psi]}. \tag{3.1}$$

$S = S_g + S_f$ is the euclidean action in the continuum (2.1),(2.4) and the functional integral extends over those fields which satisfy the boundary conditions

$$A_k(x)|_{x_0=0} = C_k^\Lambda(\mathbf{x}), \qquad\qquad A_k(x)|_{x_0=T} = C'_k(\mathbf{x}),$$
$$P_+\psi(x)|_{x_0=0} = \Lambda(\mathbf{x})\rho_+(\mathbf{x}), \qquad P_-\psi(x)|_{x_0=T} = \rho'_-(\mathbf{x}), \tag{3.2}$$
$$\bar{\psi}(x)P_-|_{x_0=0} = \bar{\rho}_-(\mathbf{x})\Lambda(\mathbf{x})^{-1}, \qquad \bar{\psi}(x)P_+|_{x_0=T} = \bar{\rho}'_+(\mathbf{x}).$$

Here, $C^\Lambda$ denotes the gauge transform of the boundary gauge field $C$,

$$C_k^\Lambda = \Lambda C_k \Lambda^{-1} + \Lambda \partial_k \Lambda^{-1}, \tag{3.3}$$

with the time independent gauge function $\Lambda(\mathbf{x})$. The appearance of the gauge transformed boundary fields in eq.(3.2), together with the integration over the gauge function $\Lambda(\mathbf{x})$ corresponds to a projection on the physical subspace of gauge invariant wave functions. On the lattice, this projection is equivalent to the integration over the timelike link variables [4].

All fields are periodic in space. To preserve periodicity under a gauge transformation, only periodic gauge functions are admitted. In particular $\Lambda(\mathbf{x})$ has to be periodic and can thus be interpreted as a mapping from the 3 dimensional torus to the group $SU(N)$. Such functions fall into disconnected topological classes which are labeled by an integer winding number

$$n = \frac{1}{24\pi^2} \int_0^L d^3\mathbf{x}\, \epsilon_{klj}\, \text{tr}\left\{ \left(\Lambda \partial_k \Lambda^{-1}\right)\left(\Lambda \partial_l \Lambda^{-1}\right)\left(\Lambda \partial_j \Lambda^{-1}\right) \right\}. \tag{3.4}$$

It is possible to convert the integral over $\Lambda$ to a sum over winding numbers by making use of gauge invariance [3]. To this end fixed representatives $\Lambda_n$ are chosen for gauge functions with winding number $n$. The Schrödinger functional then reads, up to a field independent normalization factor,

$$\mathcal{Z}[\bar{\rho}'_+, \rho'_-, C'; \bar{\rho}_-, \rho_+, C] = \sum_{n=-\infty}^{\infty} \int D[\psi]D[\bar{\psi}]D[A]\, e^{-S[A, \bar{\psi}, \psi]}. \tag{3.5}$$

For given $n$, the boundary conditions (3.2) are to be taken with the representative $\Lambda_n$ instead of $\Lambda$ and for later convenience we make the choice $\Lambda_0 = 1$.

*Minima of the action and gauge group*

At small values of the gauge coupling $g_0$, the dominant contributions to the functional integral are expected from small neighborhoods around the absolute minima of the action. We make the assumption that the minima of the action are determined by the pure gauge theory alone. In this respect quark fields only play a secondary rôle.

The absolute minima of the pure gauge action depend on the choice of boundary gauge fields. In the following, we consider only those boundary fields for which the minimizing configuration is unique (up to gauge transformations) and attained in the winding number $n = 0$ sector. The existence of such boundary fields has been established in ref. [3]. It is the corresponding minimal action configuration $B_\mu(x)$ which is referred to as the induced background gauge field.

With these assumptions, the gauge group is essentially determined as the group $\hat{\mathcal{G}}$ of gauge transformations which leave the boundary gauge fields intact. In order to further specify $\hat{\mathcal{G}}$ we assume irreducible boundary gauge fields, i.e. any gauge function $\Lambda(\mathbf{x})$ for which $C^\Lambda = C$ must be constant and proportional to the identity. A SU($N$) gauge transformation $\Omega \in \hat{\mathcal{G}}$ hence satisfies

$$\Omega(x) = \begin{cases} z_m & \text{at } x_0 = 0, \\ z_{m'} & \text{at } x_0 = T, \end{cases} \tag{3.6}$$

with $z_m = \exp 2\pi i m/N$, and some integer numbers $m$ and $m'$. Since constant gauge transformations act trivially on gauge fields, the group which is relevant for the gauge fixing procedure is $\mathcal{G} = \hat{\mathcal{G}}/\mathcal{Z}_N$. Here, $\mathcal{Z}_N$ denotes the center of SU($N$) which consists of all elements $z_m$, $m = 0, .., N-1$. $\mathcal{G}$ can thus be identified with the $m' = 0$ component of $\hat{\mathcal{G}}$. This group acts freely on gauge fields and has $N$ disconnected components labeled by $m$. If we choose fixed representatives $\Omega_m$ for each component $\mathcal{G}_m$ of $\mathcal{G}$, each element $g \in \mathcal{G}_m$ can be uniquely represented as

$$g = \Omega_m h, \tag{3.7}$$

where $h$ is an element of $\mathcal{G}_0$, the component containing the identity.

*Degeneracy of the saddle point*

In perturbation theory, one chooses a point on the gauge orbit of the minimal action configuration $B$ and parametrizes the *infinitesimal* fluctuations around this minimum, $A = B + g_0 q$. The gauge fixing procedure then separates the infinitesimal gauge directions from the physical field fluctuations and hence defines a saddle point.

Since the gauge group $\mathcal{G}$ consists of $N$ disconnected components, any gauge orbit decomposes accordingly. More precisely, the gauge orbit of a gauge potential $A$ is the union of the orbits of all $A^{\Omega_m}$ under the identity component $\mathcal{G}_0$. Hence, the gauge fixing procedure selects a saddle point on each component of the gauge orbit. This does not cause any problem in the pure gauge theory, because each component makes the same contribution to the functional integral, i.e. the saddle point has an $N$-fold degeneracy. However, when quark fields with non-vanishing boundary fields are included, the degeneracy is lifted, because the boundary conditions for the quark fields are not the same on the different components of a gauge orbit.

A nice method to deal with this situation starts with the observation, that a gauge transformation with $\Omega_m$ leads to

$$\mathcal{Z}[\bar{\rho}'_+, \rho'_-, C'; \bar{\rho}_-, \rho_+, C] = \mathcal{Z}[\bar{\rho}'_+, \rho'_-, C'; \bar{\rho}_- z_m^{-1}, z_m \rho_+, C]. \tag{3.8}$$

One may take the average over all values of $m$,

$$\mathcal{Z} = \frac{1}{N} \sum_{m=0}^{N-1} \mathcal{Z}[\bar{\rho}'_+, \rho'_-, C'; \bar{\rho}_- z_m^{-1}, z_m \rho_+, C]. \tag{3.9}$$

Next, we shift the fermionic integration variables in order to integrate over fields satisfying homogeneous boundary conditions. The dependence on the quark boundary fields and the boundary values $z_m$ of the gauge functions now resides in the action and is displayed by a superscript $(m)$,

$$\mathcal{Z}[\bar{\rho}'_+, \rho'_-, C'; \bar{\rho}_-, \rho_+, C] = \frac{1}{N} \sum_{m=0}^{N-1} \int \mathrm{D}[A]\mathrm{D}[v]\mathrm{D}[\bar{v}] e^{-S^{(m)}[A, \bar{\chi}+\bar{v}, \chi+v]}. \tag{3.10}$$

If the boundary values $z_m$ are interpreted as additional dynamical variables of the action, transforming as

$$z_m \to z_{l+m} \tag{3.11}$$

under a gauge transformation $\Omega_l$, the enlarged action is gauge invariant, i.e.

$$S^{(m+l)}[A^{\Omega_l}, \bar{\chi}\Omega_l^{-1}, \Omega_l \chi] = S^{(m)}[A, \bar{\chi}, \chi]. \tag{3.12}$$

Furthermore, the summation over $m$ can be regarded as part of the measure, so that the measure and the action are separately invariant under the action of the gauge group $\mathcal{G}$. This is the standard situation to which the gauge fixing procedure can be applied as usual. Leaving the sum over $m$ until the very end, one can treat the functional integral for fixed value of $m$ in exactly the same way as the pure gauge theory functional. In particular, one recovers the $N$-fold degeneracy of the saddle point, because the fermionic boundary conditions now transform covariantly from one component of a gauge orbit to any other one.

*Dimensional regularization*

The formulation of the dimensionally regularized theory starts with the extension of space-time to a $D$-dimensional manifold. For the $p$ additional dimensions we choose the torus $T^p = S^1 \times S^1 \times \cdots \times S^1$. Periodic boundary conditions are assumed for all fields so that the additional dimensions are analogous to the spatial ones.

As usual in dimensional regularization, external momenta and fields have physical components only. Concerning the Schrödinger functional, this means that the boundary fields are independent of the extra dimensions. Consistency then requires that also the gauge functions $\Omega \in \mathcal{G}$ and $\Lambda_n$ depend on the physical dimensions only.

Using the conventions of appendix A, the action in $D = 4 + p$ dimensions reads

$$S^{(m)}[A, \bar{\chi} + \bar{v}, \chi + v] = S_g[A] + S_f[A, \bar{v}, v] + S_f^{(m)}[A, \bar{\chi}, \chi], \qquad (3.13)$$

with

$$S_g[A] = -\frac{1}{2g_0^2} \int d^D \hat{x}\, \mathrm{tr}\,\{F_{\hat{\mu}\hat{\nu}}(\hat{x}) F_{\hat{\mu}\hat{\nu}}(\hat{x})\},$$

$$S_f[A, \bar{v}, v] = \int d^D \hat{x}\, \bar{v}(\hat{x}) \Big(\gamma_{\hat{\mu}} \partial_{\hat{\mu}} + \gamma_{\hat{\mu}} A_{\hat{\mu}}(\hat{x}) + m\Big) v(\hat{x}), \qquad (3.14)$$

$$S_f^{(m)}[A, \bar{\chi}, \chi] = -\int d^{D-1} \hat{\mathbf{x}}\, \big[\bar{\chi}(\hat{x}) P_- \chi(\hat{x})\big]_{x_0=0}$$

$$\qquad\qquad - \int d^{D-1} \hat{\mathbf{x}}\, \big[\bar{\chi}(\hat{x}) P_+ \chi(\hat{x})\big]_{x_0=T}.$$

The fields $\chi$ and $\bar{\chi}$ are solutions of the equations of motion which follow from $S^{(m)}$. One finds

$$\chi(\hat{x}) = \int d^{D-1} \hat{\mathbf{x}}'\, \big[\hat{S}'(\hat{x}; 0, \hat{\mathbf{x}}') z_m \rho_+(\mathbf{x}') + \hat{S}'(\hat{x}; T, \hat{\mathbf{x}}') \rho'_-(\mathbf{x}')\big], \qquad (3.15)$$

and a similar expression for $\bar{\chi}$. The $D$-dimensional quark propagator $\hat{S}'(x, x')$ satisfies

$$\big[\hat{\slashed{\partial}} + \gamma_{\hat{\mu}} A_{\hat{\mu}}(\hat{x}) + m\big] \hat{S}'(\hat{x}, \hat{x}') = \delta(\hat{x} - \hat{x}'). \qquad (3.16)$$

For the action, one derives the expression [cf.eq.(2.10)]

$$\begin{aligned}
S_f^{(m)}[A, \bar{\chi}, \chi] = -\int d^{D-1}\hat{\mathbf{x}}\, d^{D-1}\hat{\mathbf{x}}' \Big[ &\bar{\rho}_-(\mathbf{x}) \hat{S}'(0, \hat{\mathbf{x}}; 0, \hat{\mathbf{x}}') \rho_+(\mathbf{x}') \\
& + \bar{\rho}'_+(\mathbf{x}) \hat{S}'(T, \hat{\mathbf{x}}; T, \hat{\mathbf{x}}') \rho'_-(\mathbf{x}') \\
& + z_m^{-1}\, \bar{\rho}_-(\mathbf{x}) \hat{S}'(0, \hat{\mathbf{x}}; T, \hat{\mathbf{x}}') \rho'_-(\mathbf{x}') \\
& + z_m\, \bar{\rho}'_+(\mathbf{x}) \hat{S}'(T, \hat{\mathbf{x}}; 0, \hat{\mathbf{x}}') \rho_+(\mathbf{x}') \Big],
\end{aligned} \qquad (3.17)$$

where the dependence on $z_m$ is explicit. Taking into account the behavior of the propagator under a gauge transformation $\Omega$,

$$\hat{S}'(\hat{x},\hat{x}') \to \Omega(x)\hat{S}'(\hat{x},\hat{x}')\Omega(x')^{-1}, \tag{3.18}$$

one easily verifies that indeed both parts of the quark action in eq.(3.14) are invariant under the action of the gauge group $\mathcal{G}$.

*Gauge fixing*

The gauge fixing procedure may entirely be taken over from refs.[3,7]. Due to the assumptions on the minimal action configuration, only the winding number $n = 0$ sector contributes to the saddle point expansion. Restricting our attention to this sector, the gauge field $A$ is decomposed as follows,

$$A_\mu(\hat{x}) = B_\mu(x) + g_0 q_\mu(\hat{x}), \qquad A_{\bar{\mu}}(\hat{x}) = g_0 q_{\bar{\mu}}(\hat{x}). \tag{3.19}$$

The spatial components of the fluctuation field $q$ hence satisfy homogeneous boundary conditions,

$$q_{\hat{k}}(0,\hat{\mathbf{x}}) = 0, \qquad q_{\hat{k}}(T,\hat{\mathbf{x}}) = 0, \tag{3.20}$$

whereas the time component $q_0$ remains unrestricted at this point.

The so-called background gauge is implemented by choosing the gauge fixing term (with bare gauge fixing parameter $\lambda_0$) and the Faddeev-Popov action as in ref.[3]. Denoting the covariant derivative in the adjoint representation by $\nabla_{\hat{\mu}} = \partial_{\hat{\mu}} + \operatorname{Ad} B_{\hat{\mu}}$, we have

$$\begin{aligned} S_{\text{gf}}[B,q] &= -\lambda_0 \int \mathrm{d}^D \hat{x}\, \operatorname{tr}\{\nabla_{\hat{\mu}} q_{\hat{\mu}} \nabla_{\hat{\nu}} q_{\hat{\nu}}\}, \\ S_{\text{FP}}[B,q,\bar{c},c] &= 2 \int \mathrm{d}^D \hat{x}\, \operatorname{tr}\{\bar{c}\,\nabla_{\hat{\mu}}(\nabla_{\hat{\mu}} + g_0 \operatorname{Ad} q_{\hat{\mu}})c\}. \end{aligned} \tag{3.21}$$

The total gauge fixed action reads

$$\begin{aligned} S_{\text{total}}^{(m)}[B,q,\bar{c},c,\bar{\chi}+\bar{v},\chi+v] =\,& S_g[B+g_0 q] + S_{\text{gf}}[B,q] + S_{\text{FP}}[B,q,\bar{c},c] \\ & + S_f[B+g_0 q, \bar{v}, v] + S_f^{(m)}[B+g_0 q, \bar{\chi}, \chi]. \end{aligned} \tag{3.22}$$

A careful analysis shows that the Faddeev Popov ghosts and the time component of the quantum field $q$ satisfy the boundary conditions [3],

$$\nabla_0 q_0(\hat{x}) = c(\hat{x}) = \bar{c}(\hat{x}) = 0 \qquad \text{at} \quad x_0 = 0 \quad \text{and} \quad x_0 = T. \tag{3.23}$$

Finally, the gauge fixed, dimensionally regularized Schrödinger functional is given by

$$\mathcal{Z}[\bar{\rho}'_+, \rho'_-, C'; \bar{\rho}_-, \rho_+, C] = \frac{1}{N} \sum_{m=0}^{N-1} \int \mathrm{D}[q] \int \mathrm{D}[c]\mathrm{D}[\bar{c}] \int \mathrm{D}[v]\mathrm{D}[\bar{v}] \qquad (3.24)$$
$$\times\, \mathrm{e}^{-S_{\mathrm{total}}^{(m)}[B, q, \bar{c}, c, \bar{\chi}+\bar{v}, \chi+v]},$$

and provides the starting point for the saddle point expansion.

*The rôle of $z_m$*

Before turning to the saddle point expansion let us understand the rôle of the variables $z_m$. The summation over $m$ may actually be carried out if one expands the exponential of the $z_m$-dependent part of the action, viz,

$$\exp\{-S_f^{(m)}[B+g_0 q, \bar{\chi}, \chi]\} = \sum_{k=0}^{\infty} \frac{(-1)^k}{k!} \{S_f^{(m)}[B+g_0 q, \bar{\chi}, \chi]\}^k. \qquad (3.25)$$

In view of eq.(3.17), this corresponds to an expansion of the Schrödinger functional in powers of the boundary quark fields.

The $k=0$ contribution is independent of $z_m$ so that the sum over $m$ is trivial. For $k=1$, one sees that due to

$$\sum_{m=0}^{N-1} z_m = 0, \qquad (3.26)$$

only the part which contains either the fields at $x_0 = 0$ or the fields at $x_0 = T$ survives the summation over $m$. Concerning a general term in the expansion, one still obtains the contributions which involve only boundary fields at one of the boundaries. In addition, one may have pairings $z_m z_m^{-1} = 1$, and powers thereof. If we use the relation

$$\frac{1}{N} \sum_{m=0}^{N-1} z_m^l = \begin{cases} 1, & \text{if } l = nN,\ n = 0, 1, 2 \ldots \\ 0, & \text{otherwise}, \end{cases} \qquad (3.27)$$

we also see that terms which contain $z_m^N = 1$ or powers thereof contribute to the functional integral.

The physical interpretation is obvious. Only gauge invariant ("colorless") states are propagated in euclidean time. These are states, formed from the boundary quark fields, which have the quantum numbers of either "mesons" or "baryons" or are combinations of both. This physical picture is thus seen to be a consequence of gauge invariance.

## 4. The saddle point expansion

Having understood the rôle of the variables $z_m$, one may leave the summation over $m$ until the very end and restrict attention to the gauge fixed functional integral with fixed value of $m$,

$$\mathcal{Z}^{(m)}[\bar{\rho}'_+, \rho'_-, C'; \bar{\rho}_-, \rho_+, C] \stackrel{\text{def}}{=} \int \text{D}[q] \int \text{D}[c]\text{D}[\bar{c}] \int \text{D}[v]\text{D}[\bar{v}] \\ \times e^{-S^{(m)}_{\text{total}}[B, q, \bar{c}, c, \bar{\chi}+\bar{v}, \chi+v]}. \tag{4.1}$$

From the technical point of view, the evaluation of the $\mathcal{Z}^{(m)}$ is the same for all values of $m$. In what follows we will therefore restrict attention to $\mathcal{Z}^{(0)}$. Eventually, all other contributions may be obtained by appropriately replacing the boundary quark fields.

It proves convenient to define an auxiliary quantity $\Gamma$ through

$$\Gamma[B, \bar{\psi}_{cl}, \psi_{cl}] \stackrel{\text{def}}{=} -\ln \mathcal{Z}^{(0)}[\bar{\rho}'_+, \rho'_-, C'; \bar{\rho}_-, \rho_+, C]. \tag{4.2}$$

By slight abuse of language, $\Gamma[B, \bar{\psi}_{cl}, \psi_{cl}]$ will be called the effective action in the following. Its arguments are the classical fields of sect.2, with $A_\mu$ replaced by the background field $B_\mu$. Due to the relation

$$\chi(\hat{x}) = \psi_{cl}(x) - g_0 \int \text{d}^D \hat{x}_1 \hat{S}'(\hat{x}, \hat{x}_1) \slashed{q}(\hat{x}_1)\psi_{cl}(x_1), \tag{4.3}$$

and an analogous one for $\bar{\chi}$, the classical fields are equal to $\chi$ and $\bar{\chi}$ for vanishing gauge coupling $g_0$ and $z_m = 1$. Using eq.(4.3), one may rewrite the action as a function of the classical fields, viz

$$S^{(0)}_f[B+g_0 q, \bar{\chi}, \chi] = S^{(0)}_f[B+g_0 q, \bar{\psi}_{cl}, \psi_{cl}] \\ + g_0 \int \text{d}^D \hat{x}\, \bar{\psi}_{cl}(x)\slashed{q}(\hat{x})\psi_{cl}(x) \\ - g_0^2 \int \text{d}^D \hat{x}_1 \text{d}^D \hat{x}_2\, \bar{\psi}_{cl}(x_1)\slashed{q}(\hat{x}_1)\hat{S}'(\hat{x}_1, \hat{x}_2)\slashed{q}(\hat{x}_2)\psi_{cl}(x_2). \tag{4.4}$$

The interesting property of the effective action is its field dependence. Henceforth, we will drop field independent additive contributions without further notice. This corresponds to the omission of field independent multiplicative contributions to $\mathcal{Z}^{(0)}$. At the end, one therefore has to make sure that all components $\mathcal{Z}^{(m)}$ are normalized in the same way. This can be achieved, e.g. by referring to their common value in the case of vanishing quark boundary fields.

*Background field gauge symmetry*

The effective action is invariant under *arbitrary* gauge transformations of the background field and the classical quark fields,

$$\Gamma[B^\Omega, \bar{\psi}_{cl}\Omega^{-1}, \Omega\psi_{cl}] = \Gamma[B, \bar{\psi}_{cl}, \psi_{cl}]. \tag{4.5}$$

To see this explicitly, first note that the total gauge fixed action (3.22) is invariant under the combination of a gauge transformation $\Omega$ for the background fields and covariant rotation of the fluctuation fields,

$$\begin{aligned}
q_{\hat{\mu}}(\hat{x}) &\to \Omega(x)\, q_{\hat{\mu}}(\hat{x})\, \Omega(x)^{-1}, \\
\bar{c}(x) &\to \Omega(x)\, \bar{c}(\hat{x})\, \Omega(x)^{-1}, \\
c(\hat{x}) &\to \Omega(x)\, c(\hat{x})\, \Omega(x)^{-1}, \\
\bar{v}(\hat{x}) &\to \bar{v}(\hat{x})\, \Omega^{-1}, \\
v(\hat{x}) &\to \Omega(x)\, v(\hat{x}).
\end{aligned} \tag{4.6}$$

The covariant rotation corresponds to a unitary transformation in the spaces of fields integrated over. The functional measure is left invariant and the gauge invariance of the effective action hence follows. In particular, it should be emphasized that the effective action is independent of the gauge fixing parameter $\lambda_0$. One is therefore free to set $\lambda_0 = 1$ which is the most convenient choice for the perturbative calculations to be carried out later.

*Expansion in powers of $g_0$*

The effective action has an expansion in powers of the coupling constant $g_0$,

$$\Gamma = g_0^{-2}\Gamma_0 + \Gamma_1 + g_0^2 \Gamma_2 + \mathrm{O}(g_0^4). \tag{4.7}$$

To determine the coefficients of this expansion, we first expand the action in powers of the bare coupling. To O(1), the action reads

$$\begin{aligned}
S^{(0)}_{\mathrm{total}}[B, q, \bar{c}, c, \bar{\chi}+\bar{v}, \chi+v] &= S_g[B] + S^{(0)}_f[B, \psi_{cl}, \psi_{cl}] + \int \mathrm{d}^D \hat{x}\, \bar{v}\left(\slashed{D}+m\right) v \\
&\quad - 2\int \mathrm{d}^D \hat{x}\, \mathrm{tr}\left\{\tfrac{1}{2}q_\mu\left(\hat{\Delta}_1\, q\right)_\mu + \bar{c}\,\hat{\Delta}_0\, c\right\} + \mathrm{O}(g_0).
\end{aligned} \tag{4.8}$$

Denoting the field strength tensor associated to the background gauge field $B_\mu$ by $G_{\mu\nu}$, the fluctuation operators for the gluon and ghost fields are defined through

$$\begin{aligned}
(\hat{\Delta}_1 q)_{\hat{\mu}} &= -\nabla_{\hat{\nu}}\nabla_{\hat{\nu}} q_{\hat{\mu}} + (1-\lambda_0)\nabla_{\hat{\mu}}\nabla_{\hat{\nu}} q_{\hat{\nu}} - 2[G_{\hat{\mu}\hat{\nu}}, q_{\hat{\nu}}], \\
\hat{\Delta}_0 c &= -\nabla_{\hat{\nu}}\nabla_{\hat{\nu}} c.
\end{aligned} \tag{4.9}$$

The first order term of the effective action is the gauge part of the classical action,

$$\Gamma_0[B, \bar{\psi}_{cl}, \psi_{cl}] \equiv \Gamma_0[B] = g_0^2 \, S_g[B] \qquad (4.10)$$

The next order term $\Gamma_1$ receives contributions from the determinants of the fluctuation operators for ghost, gluon and quark fields. To have a unified notation, we define the operator $\hat{\Delta}_2$,

$$\hat{\Delta}_2 \stackrel{\text{def}}{=} (\hat{\slashed{D}} + m)(-\hat{\slashed{D}} + m), \qquad (4.11)$$

with the covariant derivative $D_{\hat{\mu}} = \partial_{\hat{\mu}} + B_{\hat{\mu}}$. $\hat{\Delta}_2$ is defined on spinors $v(\hat{x})$ which satisfy the boundary conditions

$$\begin{aligned} P_- v(x)|_{x_0=0} &= 0, & (D_0 - m) P_+ v(x)|_{x_0=0} &= 0, \\ P_+ v(x)|_{x_0=T} &= 0, & (D_0 + m) P_- v(x)|_{x_0=T} &= 0. \end{aligned} \qquad (4.12)$$

It is not difficult to verify that the effective action to this order is formally given by

$$\Gamma_1[B, \bar{\psi}_{cl}, \psi_{cl}] = S_f^{(0)}[B, \bar{\psi}_{cl}, \psi_{cl}] + \tfrac{1}{2} \ln \det \hat{\Delta}_1 - \ln \det \hat{\Delta}_0 - \tfrac{1}{2} \ln \det \hat{\Delta}_2. \qquad (4.13)$$

*Evaluation of the determinants*

The operators $\hat{\Delta}_i$ ($i = 0, 1, 2$) are symmetric, elliptic (for $\lambda_0 > 0$) and bounded from below. $\hat{\Delta}_0$ is positive and our assumptions on the background field are such that this is the case for $\hat{\Delta}_1$ and $\hat{\Delta}_2$ as well. All three operators have a complete set of smooth eigenfunctions and a discrete spectrum of positive eigenvalues. Their heat kernels (cf. appendix B) are therefore well-defined and can be used to define the logarithm of the determinants as meromorphic functions of $p = -2\varepsilon$. For the pure gauge theory, this has been carried out in ref.[3]. Neglecting terms which vanish with $\varepsilon$, the result is

$$\begin{aligned} \ln \det \hat{\Delta}_0 &= -\left[ \frac{1}{\varepsilon} + \ln 4\pi - \gamma_{\text{E}} \right] L^{-2\varepsilon} \alpha_0(\Delta_i) - \zeta'(0|\Delta_0), \\ \ln \det \hat{\Delta}_1 &= -\left[ \frac{1}{\varepsilon} + \ln 4\pi - \gamma_{\text{E}} \right] L^{-2\varepsilon} \alpha_0(\Delta_1) + 2 L^{-2\varepsilon} \alpha_0(\Delta_0) - \zeta'(0|\Delta_1). \end{aligned} \qquad (4.14)$$

The derivative of the zeta function $\zeta(s|\Delta)$ at zero is given by

$$\zeta'(0|\Delta) = \alpha_0(\Delta)\gamma_{\text{E}} + \int_0^\infty \frac{dt}{t} \left[ \text{Tr}\, e^{-t\Delta} - \sum_{j=0}^4 t^{-j/2} \alpha_{j/2}(\Delta) \right. \\ \left. + \theta(t-1)\alpha_0(\Delta) \right]. \qquad (4.15)$$

The coefficients $\alpha_{j/2}(\Delta)$ appear in the Seeley-DeWitt expansion for the trace of the heat kernels [8,9], viz

$$\mathrm{Tr}\, e^{-t\Delta} \stackrel{t\to 0}{\sim} \alpha_2(\Delta)\, t^{-2} + \alpha_{3/2}(\Delta)\, t^{-3/2} + \alpha_1(\Delta)\, t^{-1} + \ldots. \qquad (4.16)$$

Since $\hat{\Delta}_2$ has the same properties as $\hat{\Delta}_0$, we need not discuss it separately and infer its contribution from eq.(4.14). In order to obtain the singular part of the effective action it thus remains to determine the Seeley coefficient $\alpha_0$ for the 3 fluctuation operators.

The Seeley coefficients can be calculated using known techniques [10,11]. They get two kinds of contributions. The so-called volume terms contribute only for even $j$ and are given as local polynomials in the gauge field $B_\mu$ and its derivatives, integrated over the volume. The boundary terms may contribute to all terms except the leading one ($j = 4$). In contrast to the volume terms, the local composite fields have support only at $x_0 = 0$ or at $x_0 = T$ and are integrated over the corresponding hyper-planes. On dimensional grounds, the canonical dimension of the local composite fields must be $4 - j$ for the volume terms, and $3 - j$ for the boundary terms.

It is at this point that one can take full advantage of the fact that the background gauge does not violate the gauge symmetry in the background field. Therefore, the coefficients which appear in the Seeley-DeWitt expansion are gauge invariant. On the other hand, a boundary contribution to $\alpha_0(\Delta)$ must arise from a dimension 3 polynomial in the background field and its derivatives. Since such a gauge invariant polynomial (with even parity) does not exist, one may conclude that $\alpha_0$ only receives the usual contribution from the volume term.

Using the known result (see e.g. [12]), one gets

$$\begin{aligned}
\alpha_0(\Delta_0) &= \frac{N}{96\pi^2} \int d^4x\, \mathrm{tr}\{G_{\mu\nu} G_{\mu\nu}\}, \\
\alpha_0(\Delta_1) &= -20\, \alpha_0(\Delta_0), \\
\alpha_0(\Delta_2) &= -\frac{4n_f}{N}\, \alpha_0(\Delta_0).
\end{aligned} \qquad (4.17)$$

*Renormalization of the coupling constant*

Using the result (4.17), we notice that the divergent part of the effective action is proportional to the classical action of the background gauge field,

$$\Gamma_1[B, \bar\psi_{cl}, \psi_{cl}] \stackrel{\varepsilon \to 0}{=} -\frac{1}{3\varepsilon}\frac{11N - 2n_f}{16\pi^2} \Gamma_0[B] + \mathrm{O}(1). \qquad (4.18)$$

The singularity will be canceled by the usual coupling constant renormalization. To see this, let $\mu$ be the normalization mass and define the renormalized coupling $g_{\mathrm{MS}}$ in the minimal scheme (MS) of dimensional regularization as usual through

$$g_0^2 = \mu^{2\varepsilon} g_{\text{MS}}^2 \left\{ 1 + \sum_{l=1}^{\infty} z_l(\varepsilon) \, g_{\text{MS}}^{2l} \right\}. \tag{4.19}$$

According to the convention of the minimal scheme, the singular coefficients contain only the poles in $\varepsilon$ without constant part, i.e.,

$$z_l(\varepsilon) = \sum_{k=1}^{l} z_{lk} \, \varepsilon^{-k}. \tag{4.20}$$

The effective action near four dimensions as a function of the renormalized coupling is then given by

$$\Gamma[B, \bar{\psi}_{cl}, \psi_{cl}] = \mu^{-2\varepsilon} \left[ \frac{1}{g_{\text{MS}}^2} - z_1(\varepsilon) \right] \Gamma_0[B] + \Gamma_1[B, \bar{\psi}_{cl}, \psi_{cl}] + \text{O}(g_{\text{MS}}^2). \tag{4.21}$$

The effective action is hence finite up to terms of the order $\text{O}(g_{\text{MS}}^2)$, provided one sets

$$z_1(\varepsilon) = -\frac{1}{3\varepsilon} \frac{11N - 2n_f}{16\pi^2}, \tag{4.22}$$

which is the same coefficient as usual [13–15]. In 4 dimensions, the field dependent parts of the effective action to this order are then given by

$$\begin{aligned}\Gamma[B, \bar{\psi}_{cl}, \psi_{cl}]\Big|_{D=4} &= \left\{ \frac{1}{g_{\text{MS}}^2} - \frac{11N - 2n_f}{48\pi^2} [\ln 4\pi\mu^2 - \gamma_{\text{E}}] - \frac{N}{48\pi^2} \right\} \Gamma_0[B] \\ &\quad - \tfrac{1}{2}\zeta'(0|\Delta_1) + \zeta'(0|\Delta_0) + \tfrac{1}{2}\zeta'(0|\Delta_2) \\ &\quad + S_f^{(0)}[B, \bar{\psi}_{cl}, \psi_{cl}] + \text{O}(g_{\text{MS}}^2).\end{aligned} \tag{4.23}$$

## 5. One-loop boundary effects

In order to observe the effect of the boundary one has to look for one-loop graphs involving the quark boundary fields, or, equivalently, the classical quark fields. It turns out that only 1 graph, the quark self-energy needs to be calculated explicitly. Its evaluation is done using a method which is originally due to Lüscher [12] and may be adapted to the case at hand.

*Preliminaries*

The effective action is a functional of the classical quark fields which play the rôle of external sources. Expanding the effective action in powers of the classical fields, one may define $n$-point functions as the coefficients of this series. In order to identify possible boundary counterterms recall that these must be given by local composite fields of canonical dimension 3. Therefore, it is sufficient to consider only the 2-point function. The corresponding contribution to the effective action should reflect the form of the boundary counterterms, which are expected to cancel the boundary divergences in the higher $n$-point functions as well.

To identify the graphs which contribute to the 2-point function, we inspect the interaction part of the gauge fixed action (3.22). Taking into account the couplings to the classical quark fields (4.4), one can construct 2 types of one-loop graphs which contribute to the 2-point function. First, a tadpole graph with a ghost, gluon or quark field in the loop is obtained, due to the coupling of the classical quark fields to a single gluon. We will argue that no divergent part is to be expected from the tadpole graphs. In order to present the argument, it is however useful to first acquire some experience. This will be done by considering the second contribution at this order, which is the self-energy graph for the quark field,

$$\Gamma_2[B, \bar{\psi}_{cl}, \psi_{cl}] = -g_0^2 \int \mathrm{d}^D \hat{x}_1 \mathrm{d}^D \hat{x}_2\, \bar{\psi}_{cl}(x_1)\hat{\Sigma}(\hat{x}_1, \hat{x}_2)\psi_{cl}(x_2) + \ldots . \tag{5.1}$$

Explicitly, the one-loop expression for the self-energy $\hat{\Sigma}$ is given by

$$\hat{\Sigma}(\hat{x}, \hat{x}') = \hat{D}(\hat{x}, \hat{x}')^{ab}_{\hat{\mu}\hat{\nu}}\gamma_{\hat{\mu}}T^a \hat{S}(\hat{x}, \hat{x}')\gamma_{\hat{\nu}}T^b. \tag{5.2}$$

Here, $\hat{D}(\hat{x}, \hat{x}')^{ab}_{\hat{\mu}\hat{\nu}}$ is the $D$-dimensional free gluon propagator. It is related to the heat kernel through

$$\hat{D}(\hat{x}, \hat{x}')^{ab}_{\hat{\mu}\hat{\nu}} = \int_0^\infty \mathrm{d}t\, \hat{K}_t(\hat{x}, \hat{x}'|\hat{\Delta}_1)^{ab}_{\hat{\mu}\hat{\nu}}. \tag{5.3}$$

To have a similar representation of the quark propagator it is convenient to define the integral kernel

$$\hat{\mathcal{K}}_t(\hat{x}, \hat{x}'|\hat{\Delta}_2) = (-\hat{\slashed{D}} + m)\hat{K}_t(\hat{x}, \hat{x}'|\hat{\Delta}_2), \tag{5.4}$$

which is defined such that

$$\hat{S}(\hat{x}, \hat{x}') = \int_0^\infty \mathrm{d}t\, \hat{\mathcal{K}}_t(\hat{x}, \hat{x}'|\hat{\Delta}_2). \tag{5.5}$$

In the following, the divergent part of the quark self-energy will be evaluated at one-loop order.

*The method*

The standard techniques of dimensional regularization [16–18] work in momentum space and thus rely on translation invariance. The latter is broken by the presence of the boundary as well as by the background fields. We therefore follow the method proposed in ref.[12], which may be adapted to the case of manifolds with a boundary.

The procedure is briefly described as follows. One first writes down the expression for a given Feynman graph in position space according to the appropriate Feynman rules. Next, one inserts the heat kernel representation of all propagators and performs a Fourier transformation in the $p$ extra directions. Taking into account translation invariance and the vanishing of external momenta in these directions, the $p$-dimensional integrals are either trivial due to $\delta$-functions, or gaussian and can be calculated explicitly. The outcome is an integral over the proper times with an integrand consisting of the purely 4-dimensional heat kernels, multiplied by an explicitly known, $p$-dependent function of the proper times. The latter regulates the proper time integrations at the lower end, for suitable choice of $p = -2\varepsilon$, (usually for sufficiently large $\operatorname{Re}\varepsilon$). Smearing the resulting expression with a test function, an ordinary function of $p = -2\varepsilon$ is obtained. To analytically continue this function to $\varepsilon = 0$, one must find out for which values of the proper times the integrand is not integrable without the regulating factor. This is usually the case when some or all of the proper time parameters simultaneously approach zero. By subtracting and adding the asymptotic expansion of the integrand in these regions, one may isolate the poles in $\varepsilon$ and eventually obtain the Feynman diagram as a meromorphic function of $\varepsilon$.

*Application to the quark self-energy*

Using the notational conventions of appendix A, we define the Fourier transform in the extra dimensions,

$$\Sigma(x,x') \stackrel{\text{def}}{=} \int \mathrm{d}^p y \, \hat{\Sigma}(\hat{x},\hat{x}'). \tag{5.6}$$

The self-energy is a distribution which is to be smeared with a smooth and spatially periodic test function $f(x,x')$. At the boundary, $f$ may take any finite value and it is assumed that all its derivatives at the boundary exist. The test function has color, flavor and spinor indices which match those of the self energy so that $\Sigma(f)$, defined by

$$\Sigma(f) \stackrel{\text{def}}{=} \int \mathrm{d}^4 x \, \mathrm{d}^4 x' \operatorname{tr}\{f(x,x')\Sigma(x,x')\}, \tag{5.7}$$

is an ordinary function of $p = -2\varepsilon$. Note that the contribution to the effective action (5.1) has exactly this structure, provided the classical quark fields are considered as test functions. This will be used in sect.6, when the renormalization of the effective action will be discussed.

When the heat kernel representation of the propagators (5.3),(5.5) is inserted in

eq.(5.2), one gets

$$\hat{\Sigma}(\hat{x},\hat{x}') = \int_0^\infty dt_1 dt_2 \; \hat{K}_{t_1}(\hat{x},\hat{x}'|\hat{\Delta}_1)^{ab}_{\hat{\mu}\hat{\nu}} \gamma_{\hat{\mu}} T^a \hat{\mathcal{K}}_{t_2}(\hat{x},\hat{x}'|\hat{\Delta}_2) \gamma_{\hat{\nu}} T^b. \tag{5.8}$$

Using the factorization properties of the heat kernels (appendix B), the integration over the extra dimensions and smearing with a test function yields

$$\Sigma(f) = \int_0^\infty dt_1 dt_2 \; [r(t_1+t_2)]^p \, I(f,t_1,t_2), \tag{5.9}$$

with the explicitly known function $r(t)$,

$$r(t) = (4\pi t)^{-1/2} \sum_{n=-\infty}^{\infty} e^{-n^2 L^2/4t}, \tag{5.10}$$

and the integrand $I = I_1 + I_0$, given explicitly by

$$\begin{aligned} I_1(f,t_1,t_2) &= \int d^4x \, d^4x' \, \text{tr} \left\{ f(x,x) K_{t_1}(x,x'|\Delta_1)^{ab}_{\mu\nu} \gamma_\mu T^a \mathcal{K}_{t_2}(x,x'|\Delta_2) \gamma_\nu T^b \right\}, \\ I_0(f,t_1,t_2) &= \int d^4x \, d^4x' \, \text{tr} \left\{ f(x,x) K_{t_1}(x,x'|\Delta_0)^{ab} \gamma_{\bar{\mu}} T^a \mathcal{K}_{t_2}(x,x'|\Delta_2) \gamma_{\bar{\mu}} T^b \right\}. \end{aligned} \tag{5.11}$$

The contraction of the $p$ additional $\gamma$-matrices in $I_0(f,t_1,t_2)$ yields a factor of $p$, so that the dependence of the integrand on $p = -2\varepsilon$ is completely explicit.

*Properties of the function $I(f,t_1,t_2)$*

$I(f,t_1,t_2)$ is an ordinary function of the proper times $t_1$ and $t_2$. Due to the positivity of the fluctuation operators, $I(f,t_1,t_2)$ vanishes exponentially for large values of one of its arguments. Concerning its behavior at small values of the proper times we first recall that the heat kernels satisfy

$$\lim_{t \to 0} \int d^4x \, d^4x' \, \text{tr} \left\{ f(x,x') K_t(x,x'|\Delta) \right\} = \int d^4x \, \text{tr} \, f(x,x). \tag{5.12}$$

Note that the heat kernels themselves and their derivatives are, for non-zero proper time $t$, valid test functions. If $t_2 > 0$ is held fixed, we may therefore conclude that $I(f,0,t_2)$ exists. Using similar techniques as in appendix C, one may also show that $I(f,t_1,0)$ exists for $t_1 > 0$. Hence, if one of the proper times in the function $I(f,t_1,t_2)$ is kept different from zero, the other can be taken to zero. This yields a smooth function of the remaining proper time, which diverges when the latter approaches zero. Together with the exponential fall-off at large proper times this leads us to conclude that the function $I(f,t_1,t_2)$ is integrable everywhere, except near the origin $t_1 = t_2 = 0$.

*One-loop result*

It is convenient to perform the change of variables

$$t_1 = ts_1, \qquad t_2 = ts_2, \qquad s_1 + s_2 = 1, \tag{5.13}$$

so that the self-energy (5.9) reads

$$\Sigma(f) = \int_0^\infty \mathrm{d}t\, t\, [r(t)]^{-2\varepsilon} \int_0^1 \mathrm{d}s_1 \mathrm{d}s_2\, \delta(s_1 + s_2 - 1) I(f, ts_1, ts_2). \tag{5.14}$$

In appendix C, the asymptotic small $t$ behavior of $I(f, ts_1, ts_2)$ is determined,

$$I(f, ts_1, ts_2) \stackrel{t \to 0}{\sim} A(f, s_1, s_2)\, t^{-2} + \mathrm{O}(t^{-3/2}), \tag{5.15}$$

and an explicit expression for the coefficient function $A(f, s_1, s_2)$ is obtained.

Since the regulating function $r(t)$ (5.10) behaves asymptotically as $t^{-1/2}$, the whole integrand in eq.(5.14) behaves as $t^{\varepsilon - 1}$ at the lower end of the $t$ integration. It hence follows that the self-energy (5.14) is a priori only defined for $\mathrm{Re}\,\varepsilon > 0$. In order to analytically continue to $\varepsilon = 0$, we first split up the range of the $t$-integration at $t = 1$. The integral for $t > 1$ is analytic in $\varepsilon$. In the remaining integral we subtract and add the asymptotic expansion (5.15) of the integrand. The integral over the subtracted integrand will exist for $\mathrm{Re}\,\varepsilon > -1/2$ and the integral over the asymptotic expansion can be calculated explicitly. Near 4 dimensions, we obtain

$$\begin{aligned}\Sigma(f) \stackrel{\varepsilon \to 0}{=} &\left(\frac{1}{\varepsilon} + \ln 4\pi\right) \int_0^1 \mathrm{d}s\, A(f, s, 1-s) \\ &+ \int_0^\infty \mathrm{d}t_1 \mathrm{d}t_2 \left[I(f, t_1, t_2) - \theta(1 - t_1 - t_2)\, A(f, t_1, t_2)\right] + \mathrm{O}(\varepsilon),\end{aligned} \tag{5.16}$$

where we have used that $A(f, t_1, t_2)$ is a homogeneous function of degree $-2$ in $t_1$ and $t_2$. We will be mainly interested in the divergent part of the self-energy which is explicitly given by

$$\begin{aligned}\Sigma(f) \stackrel{\varepsilon \to 0}{=} \frac{C_\mathrm{F}}{16\pi^2 \varepsilon} &\left[-4m \int \mathrm{d}^4 x\, \mathrm{tr}\, f(x,x) \right. \\ &+ \int \mathrm{d}^4 x\, \mathrm{tr}\, \{\tfrac{1}{2}\slashed{\partial}^{(1)} f(x,x) - \tfrac{1}{2}\slashed{\partial}^{(2)} f(x,x) - \slashed{B}(x) f(x,x)\} \\ &\left. - \tfrac{3}{2} \int \mathrm{d}^3 \mathbf{x}\, \mathrm{tr}\, \{f(0, \mathbf{x}, 0, \mathbf{x}) + f(T, \mathbf{x}, T, \mathbf{x})\}\right] + \mathrm{O}(1).\end{aligned} \tag{5.17}$$

*Tadpole graphs*

Before turning to the renormalization of the effective action and the Schrödinger functional, we come back to the tadpole graphs mentioned at the beginning of this section. In these graphs, a gluon line connects the classical quark fields with a loop, in which ghost, gluon or quark fields may circulate. The contribution of all tadpole graphs to the effective actions can be parameterized by a function $T_\mu$

$$\Gamma_2[B,\bar\psi_{cl},\psi_{cl}]\big|_{\text{tadpoles}} = L^{-2\varepsilon} \int \mathrm{d}^4 x\, \bar\psi_{cl}\gamma_\mu T_\mu \psi_{cl}, \tag{5.18}$$

which may depend on the background gauge field $B$ and the quark mass $m$. Furthermore, $T_\mu$ must be gauge covariant, due to the background gauge symmetry of the effective action (cf. sect.4). Note also that $T_\mu$ need not transform as a Lorentz vector, because Lorentz invariance is violated by the presence of the boundary. Fortunately, it is not necessary to treat the tadpole graphs explicitly. For, we are interested in the divergent contributions to the effective action. Such contributions cannot arise from the tadpole graphs as the following argument shows.

The lesson that can be learned from the treatment of the self-energy graph is, that only *local* terms appear in the asymptotic expansion. Possible divergences may therefore be identified using naive power counting. When applied to the tadpole graphs, this means that a divergent contribution to the function $T_\mu$ must have canonical dimension 1. The requirement of gauge covariance for $T_\mu$ excludes terms which involve the background gauge field. There could be a term proportional to the mass or a $\delta$-function, confining the term (5.18) at the boundary.

At this point, the crucial observation is that, for vanishing background field $B$, the tadpole graphs vanish identically. The reason is that each vertex contains a $SU(N)$ generator and all propagators are, for vanishing background field, diagonal in color space. The tadpole graphs therefore become proportional to the trace of a $SU(N)$ generator in either the fundamental or the adjoint representation (see appendix A). Consequently, the tadpole graphs can only contribute finite terms to the effective action and may therefore be ignored as far as renormalization is concerned.

# 6. Renormalization

*Renormalized effective action*

All divergent one-loop contributions to the 2-point function are assembled in the self-energy (5.16). To find the corresponding contributions to the effective action, we note that the classical solutions $\psi_{cl}$ and $\bar\psi_{cl}$ have all the properties which are required for a test function (cf. sect.5). We define the special test function,

$$f_{cl}(x,x')_{st;ij;\alpha\beta} \stackrel{\text{def}}{=} \bar\psi_{cl}(x)_{si\alpha}\psi_{cl}(x')_{tj\beta}. \tag{6.1}$$

Recalling eqs.(5.1),(5.6) and (5.7), the contribution to the effective action is given by

$$\Gamma_2[B,\bar\psi_{cl},\psi_{cl}] = -L^{-2\varepsilon}\Sigma(f_{cl}) + \ldots . \tag{6.2}$$

If the test function $f_{cl}$ is inserted in the pole part of the self-energy (5.17), one obtains

$$\Sigma(f_{cl}) \stackrel{\varepsilon\to 0}{=} \frac{C_{\text{F}}}{16\pi^2\varepsilon}\Big[-3m\int \text{d}^4x\,\bar\psi_{cl}(x)\psi_{cl}(x) \\ -\tfrac{3}{2}\int \text{d}^3\mathbf{x}\big\{\big[\bar\psi_{cl}(x)\psi_{cl}(x)\big]_{x_0=0} + \big[\bar\psi_{cl}(x)\psi_{cl}(x)\big]_{x_0=T}\big\}\Big] + \text{O}(1). \tag{6.3}$$

To arrive at this equation, we have made use of the equations of motion for the classical quark fields. The pole contribution to the effective action is expected to reflect the form of the counterterms. The first pole has the form of a mass term and is indeed canceled by the quark mass renormalization. To see this, we introduce a renormalized mass $m_R$ through

$$m = Z_m m_R, \qquad Z_m = 1 + \sum_{n=1}^{\infty} Z_m^{(n)}(\varepsilon) g_{\text{MS}}^{2n}, \tag{6.4}$$

and choose the renormalization constant $Z_m$ according to the prescription of the minimal scheme. Using eq.(6.4) to replace the bare mass in the action of the classical quark fields leads to

$$S_f[B,\bar\psi_{cl},\psi_{cl}] = S_f[B,\bar\psi_{cl},\psi_{cl}]\big|_{m_R} \\ + g_{\text{MS}}^2 m_R Z_m^{(1)} \int \text{d}^4x\,\bar\psi_{cl}(x)\psi_{cl}(x) + \text{O}(g_{\text{MS}}^4). \tag{6.5}$$

Since the action of the classical fields is also part of the effective action [see eq.(4.23)], the quark mass renormalization eliminates the pole proportional to the quark mass in eq.(6.3), provided we choose

$$Z_m^{(1)}(\varepsilon) = \frac{-3C_{\text{F}}}{16\pi^2\varepsilon}. \tag{6.6}$$

This is the usual result for the renormalization of the quark mass in QCD [19,20].

The remaining divergence in eq.(6.3) is the boundary effect we have been looking for. One recognizes that it is proportional to the action of the classical quark fields [cf. eq.(2.9)]. Since the latter is bilinear in the classical quark fields, it is possible to eliminate the divergence by rescaling the classical quark fields appropriately. We define a further renormalization constant $Z_b$,

$$Z_b = 1 + \sum_{n=1}^{\infty} Z_b^{(n)}(\varepsilon) g_{\mathrm{MS}}^{2n}. \tag{6.7}$$

and set

$$Z_b^{(1)}(\varepsilon) = \frac{3C_{\mathrm{F}}}{16\pi^2 \varepsilon}. \tag{6.8}$$

With this definition, $\Gamma[B, Z_b^{-1/2} \bar{\psi}_{cl}, Z_b^{-1/2} \psi_{cl}]$ is finite to order $g_{\mathrm{MS}}^2$, provided the coupling constant and the mass are renormalized in the usual way.

Note that the renormalization constant $Z_b$ has nothing to do with the wave function renormalization of the quark fluctuation fields $\bar{v}$ and $v$. The latter is not needed at all, because it would just correspond to a rescaling of the integration variables of the Schrödinger functional. Note however, that the usual one-loop wave function renormalization constant (for $\lambda_0 = 1$) is contained in the pole part of the self-energy (5.17), and would have played a rôle if we had considered correlation functions such as the quark propagator.

*Back to the Schrödinger functional*

We are now prepared to discuss the renormalization of the Schrödinger functional. First recall that the effective action has been defined for the $m = 0$ part of the Schrödinger functional (4.2). To obtain the other contributions one first expresses the classical quark fields through the quark boundary fields (2.8). Next, one goes through all values for $m$ and multiplies the quark boundary fields with the appropriate factor $z_m$. This procedure guarantees that all parts $\mathcal{Z}^{(m)}$ of the Schrödinger functional are normalized in the same way. The physical implication of the summation over $m$ has already been discussed in sect.3. Here it is important to notice that the relation between the classical quark fields and the quark boundary fields is linear, for all values of $m$. If we introduce the renormalized quark boundary fields,

$$\begin{aligned} \rho_+ &= Z_b^{1/2} \rho_+^R, & \rho'_- &= Z_b^{1/2} \rho'^R_-, \\ \bar{\rho}_- &= Z_b^{1/2} \bar{\rho}_-^R, & \bar{\rho}'_+ &= Z_b^{1/2} \bar{\rho}'^R_+, \end{aligned} \tag{6.9}$$

we may therefore conclude that the Schrödinger functional,

$$\mathcal{Z}[\bar{\rho}'^R_+, \rho'^R_-, C'; \bar{\rho}^R_-, \rho^R_+, C] \tag{6.10}$$

is finite to one-loop order of perturbation theory, provided the mass and the coupling constant are renormalized in the usual way.

## 7. Conclusions

At one-loop order, the renormalization of the QCD Schrödinger functional works out according to Symanzik's expectation. In addition to the usual counterterms for quark mass and coupling constant renormalization, a new counterterm is needed to account for the boundary divergences. Its form coincides with the boundary terms already present in the classical action (cf. sect.2). Therefore, it should not come as a surprise that the boundary divergence can be absorbed in a multiplicative renormalization of the quark boundary fields. In particular, this means that for vanishing quark boundary fields, the Schrödinger functional is finite after the usual QCD renormalizations.

In the recent literature, the Schrödinger functional has been used for a systematic investigation of the continuum limit in the SU(2) and SU(3) Yang-Mills theory [21]. Since our result gives confidence that the Schrödinger functional is a universal amplitude, one may hope to extend these very successful studies to QCD.

I would like to thank M. Lüscher for numerous helpful discussions, and P. Weisz for reading the manuscript.

## Appendix A

*Conventions for dimensional regularization*

Lorentz indices are denoted by greek letters $\mu, \nu, \ldots$ and range from 0 to 3. They must be distinguished from the $D$-dimensional indices $\hat{\mu}, \hat{\nu}, \ldots$ which run from 0 to $D-1$. We will also use $\bar{\mu}, \bar{\nu}, \ldots$, which take the values $4, \ldots, D-1$. For the spatial directions we use latin indices $k, l, \ldots$ which range from 1 to 3 and their $D$-dimensional counterparts $\hat{k}, \hat{l}, \ldots$. We will make the convention that repeated indices are summed over, if not otherwise stated.

The D-dimensional Euclidean space-time manifold is taken to be a cylinder $[0,T] \times T^3 \times T^p$ where $p = D - 4$ is the number of extra dimensions. A point on this manifold is denoted by $\hat{x}$ with components $\hat{x}_{\hat{\mu}}$ which are further specified as follows

$$\hat{x}_\mu = x_\mu, \qquad \hat{x}_{\bar{\mu}} = y_{\bar{\mu}}. \tag{A.1}$$

The letter $y$ will be reserved for points on the additional $p$-dimensional manifold throughout the paper. Occasionally, we also use the spatial vector $\mathbf{x}$ and its $D-1$ dimensional counterpart $\hat{\mathbf{x}}$.

*$\gamma$-matrices*

The $\gamma$ matrices are hermitian and defined by

$$\{\gamma_{\hat{\mu}}, \gamma_{\hat{\nu}}\} = 2\,\delta_{\hat{\mu}\hat{\nu}}, \qquad \gamma_{\hat{\mu}}^{\dagger} = \gamma_{\hat{\mu}}, \qquad \hat{\mu}, \hat{\nu} = 0, \ldots, D-1. \tag{A.2}$$

The anti-commutation relation (A.2) is all that is needed to carry out the $D$-dimensional $\gamma$-algebra. For any Lorentz vector $a_{\hat{\mu}}$ we occasionally use the notation

$$\slashed{a} = \gamma_{\hat{\mu}} a_{\hat{\mu}}. \tag{A.3}$$

In particular we make the convention

$$\hat{\slashed{\partial}} = \gamma_{\hat{\mu}} \partial_{\hat{\mu}}, \qquad \slashed{\partial} = \gamma_{\mu} \partial_{\mu}, \tag{A.4}$$

where the definition $\partial_{\hat{\mu}} = \partial/\partial \hat{x}_{\hat{\mu}}$ has been used.

*su(N) conventions*

The Lie algebra su($N$) can be identified with the linear space of of all complex $N \times N$-matrices $X_{ij}$, $i,j = 1,\ldots N$, which are anti-hermitean and traceless. In other words, the matrix $X$ satisfies

$$X^{\dagger} = -X, \qquad \text{tr}\,\{X\} = 0. \tag{A.5}$$

As a vector space over $\mathbb{R}$, su($N$) has dimension $N^2 - 1$. An inner product is introduced through

$$(X, Y) = -2\,\text{tr}\,\{XY\}, \tag{A.6}$$

and may be used to define an orthonormal basis $T^a$, $a = 1, \ldots, N^2 - 1$. Any element $X \in$ su($N$) can be decomposed over this basis

$$X = X^a T^a, \qquad X^a = (X, T^a), \tag{A.7}$$

with real coefficients $X^a$. Due to the linear structure of the Lie algebra, it is sufficient to know the Lie bracket between any two elements of the basis

$$[T^a, T^b] = f^{abc} T^c, \tag{A.8}$$

in order to know them for any two elements of the Lie algebra. The structure constants $f^{abc}$ are real and totally anti-symmetric under permutations of the indices. They obey the relation

$$f^{acd} f^{bcd} = N \delta^{ab}, \tag{A.9}$$

which is a consequence of the completeness of the basis.

A representation of the Lie algebra is a vector space homomorphism $R$ which preserves the Lie bracket (A.8), i.e.

$$[R(T^a), R(T^b)] = f^{abc} R(T^c). \tag{A.10}$$

In the context of this paper only the fundamental and the adjoint representation appear. For the fundamental representation one simply has $R(X) = X$ while the adjoint representation is defined by $R(X) = \mathrm{Ad}\, X$. The representation space for the adjoint representation is the Lie algebra itself, and its action on an element $Y \in \mathrm{su}(N)$ is the commutator $\mathrm{AdX}(Y) = [X, Y]$. The representation matrices of the basis elements $T^a$ are then simply given in terms of the structure constants, i.e. we have

$$\left(\mathrm{Ad}\, T^a\right)^{bc} = -f^{abc}. \tag{A.11}$$

Finally, using eqs.(A.8),(A.9) and the normalization of the basis elements, one obtains

$$T^a T^b T^a = \frac{1}{2N} T^b, \qquad T^a T^a = -C_F, \tag{A.12}$$

where $C_F = (N^2 - 1)/2N$ is a constant which frequently appears in the loop expansion.

*Fields*

To enhance the readability, indices are often suppressed. However, a quark field $\psi(x)_{si\alpha}$ has flavor, color and spinor indices. Flavor indices are taken from the middle of the alphabet $(s, t, \ldots)$ and range from 1 to $n_f$. Spinor indices $(\alpha, \beta, \ldots)$ take the values $1, \ldots, 4$. Concerning the color structure, a quark field is a vector in the fundamental representation, with color index $i$ running from 1 to $N$.

The gluon field, its field strength tensor and the ghost fields are elements of the Lie algebra $\mathrm{su}(N)$ and can thus be decomposed according to eq.(A.7),

$$B_\mu = B_\mu^a T^a, \qquad G_{\mu\nu} = G_{\mu\nu}^a T^a, \qquad c = c^a T^a, \qquad \bar{c} = \bar{c}^a T^a. \tag{A.13}$$

Finally, we define the fields in the adjoint representation,

$$B(x)_\mu^{ab} = -f^{abc} B(x)_\mu^c, \qquad G(x)_{\mu\nu}^{ab} = -f^{abc} G(x)_{\mu\nu}^c. \tag{A.14}$$

## Appendix B

This appendix introduces the $D$-dimensional heat kernels for the fluctuation operators of ghost, gluon and quark fields. In the case of vanishing background gauge field, some explicit expressions are obtained.

*The heat kernels of the fluctuation operators*

The heat kernels of the $D$-dimensional fluctuation operators $\Delta_i$, $i = 0, 1, 2$, are defined as follows

$$\left(e^{-t\hat{\Delta}_0} c\right)(\hat{x})^a = \int d^D \hat{x}' \, \hat{K}_t(\hat{x}, \hat{x}'|\hat{\Delta}_0)^{ab} c(\hat{x}')^b,$$

$$\left(e^{-t\hat{\Delta}_1} q\right)(\hat{x})^a_{\hat{\mu}} = \int d^D \hat{x}' \, \hat{K}_t(\hat{x}, \hat{x}'|\hat{\Delta}_1)^{ab}_{\hat{\mu}\hat{\nu}} q(\hat{x}')^b_{\hat{\nu}}, \qquad (B.1)$$

$$\left(e^{-t\hat{\Delta}_2} v\right)(\hat{x})_{si\alpha} = \int d^D \hat{x}' \, \hat{K}_t(\hat{x}, \hat{x}'|\hat{\Delta}_2)_{si\alpha;tj\beta} v(\hat{x}')_{tj\beta}.$$

They satisfy the diffusion equation

$$\left(\frac{\partial}{\partial t} + \hat{\Delta}\right) \hat{K}_t(\hat{x}, \hat{x}'|\hat{\Delta}) = 0, \qquad (B.2)$$

and from the general theory [10] it can be inferred that they are smooth functions of $t$, $\hat{x}$ and $\hat{x}'$, as long as the proper time $t$ is positive. The limit $t \to 0$ yields,

$$\lim_{t \to 0} \hat{K}_t(\hat{x}, \hat{x}'|\hat{\Delta}) = \delta(\hat{x} - \hat{x}'). \qquad (B.3)$$

The heat kernels have simple factorization properties which allow to separate the dependence on the $p$ extra dimensions. With the convention

$$\Delta_i = \hat{\Delta}_i \big|_{D=4}, \qquad i = 0, 1, 2, \qquad (B.4)$$

one finds, for $\hat{\Delta}_0$ and $\hat{\Delta}_2$,

$$\hat{K}_t(\hat{x}, \hat{x}'|\hat{\Delta}_0)^{ab} = K_t^{T^p}(y, y') K_t(x, x'|\Delta_0)^{ab},$$

$$\hat{K}_t(\hat{x}, \hat{x}'|\hat{\Delta}_2)_{si\alpha;tj\beta} = K_t^{T^p}(y, y') K_t(x, x'|\Delta_2)_{si\alpha;tj\beta}, \qquad (B.5)$$

The kernel $K_t^{T^p}(y, y')$ is defined in eq.(B.8). Assuming $\lambda_0 = 1$, the factorization for $\hat{\Delta}_1$ looks as follows,

$$\hat{K}_t(\hat{x}, \hat{x}'|\hat{\Delta}_1)^{ab}_{\mu\nu} = K_t^{T^p}(y, y') K_t(x, x'|\Delta_1)^{ab}_{\mu\nu},$$

$$\hat{K}_t(\hat{x}, \hat{x}'|\hat{\Delta}_1)^{ab}_{\bar{\mu}\bar{\nu}} = K_t^{T^p}(y, y') K_t(x, x'|\Delta_0)^{ab} \delta_{\bar{\mu}\bar{\nu}}. \qquad (B.6)$$

All other components vanish identically.

*One-dimensional heat kernels*

Consider the one-dimensional laplacian $\Delta^{S^1}$ on the circle $S^1$ with circumference $L$. It may be represented by the second derivative operator $-d^2/dx^2$, acting on $L$-periodic functions of the real variable $x$. Its heat kernel is given as the integral kernel of $\exp -t\Delta^{S^1}$,

$$K_t(x, x'|\Delta^{S^1}) = (4\pi t)^{-1/2} \sum_{n=-\infty}^{\infty} e^{-(x-x'+nL)^2/4t}. \tag{B.7}$$

Using the notation of appendix A, the heat kernels for the laplacian on the tori $T^3$ and $T^p$ are then given by

$$K_t^{T^3}(\mathbf{x}, \mathbf{x}') = \prod_{k=1}^{3} K_t(x_k, x'_k|\Delta^{S^1}),$$

$$K_t^{T^p}(y, y') = \prod_{\tilde{\mu}=4}^{4+p} K_t(y_{\tilde{\mu}}, y'_{\tilde{\mu}}|\Delta^{S^1}). \tag{B.8}$$

Next, consider the laplacian on the interval $[0, T]$, acting on functions which satisfy Dirichlet (D) or Neumann (N) boundary conditions. We obtain the following one-dimensional heat kernels,

$$K_t^{\text{DD}}(x, x') = (4\pi t)^{-1/2} \sum_{n=-\infty}^{\infty} \left\{ e^{-\frac{1}{4t}(x-x'+2Tn)^2} - e^{-\frac{1}{4t}(x+x'+2Tn)^2} \right\},$$

$$K_t^{\text{NN}}(x, x') = (4\pi t)^{-1/2} \sum_{n=-\infty}^{\infty} \left\{ e^{-\frac{1}{4t}(x_0-x'_0+2Tn)^2} + e^{-\frac{1}{4t}(x+x'+2Tn)^2} \right\},$$

$$K_t^{\text{ND}}(x, x') = (4\pi t)^{-1/2} \sum_{n=0}^{\infty} \left\{ e^{-\frac{1}{4t}(|x-x'|+2nT)^2} - e^{-\frac{1}{4t}(|x-x'|-(2n+2)T)^2} \right.$$
$$\left. + e^{-\frac{1}{4t}(x+x'+2nT)^2} - e^{-\frac{1}{4t}(x+x'-(2n+2)T)^2} \right\}, \tag{B.9}$$

$$K_t^{\text{DN}}(x, x') = K_t^{\text{ND}}(T-x, T-x').$$

Here, the first superscript refers to the boundary condition at $x_0 = 0$, the second to $x_0 = T$. We will also need the heat kernels with modified Neumann conditions,

$$K_t^{\text{D}\check{\text{N}}}(x, x')|_{x=0} = 0, \qquad (\partial_x + m) K_t^{\text{D}\check{\text{N}}}(x, x')|_{x=T} = 0,$$
$$(\partial_x - m) K_t^{\check{\text{N}}\text{D}}(x, x')|_{x=0} = 0, \qquad K_t^{\check{\text{N}}\text{D}}(x, x')|_{x=T} = 0. \tag{B.10}$$

I did not succeed in obtaining explicit expressions for these heat kernels. However, if

$T$ is taken to infinity, one finds,

$$\lim_{T \to \infty} K_t^{\mathrm{D}\tilde{\mathrm{N}}}(x, x') = (4\pi t)^{-1/2} \left[ e^{-(x-x')^2/4t} - e^{-(x+x')^2/4t} \right]$$

$$\lim_{T \to \infty} K_t^{\tilde{\mathrm{N}}\mathrm{D}}(x_0, x_0') = (4\pi t)^{-1/2} \left[ e^{-(x-x')^2/4t} \right. \tag{B.11}$$

$$\left. + e^{-(x+x')^2/4t} \{ 1 - 4mt\, F(t, x+x') \} \right].$$

The parameter integral $F(t, x)$,

$$F(t, x) \stackrel{\mathrm{def}}{=} \int_0^\infty \mathrm{d}s\, e^{-t(s^2 + 2ms) - sx} \tag{B.12}$$

cannot be further simplified. For, using e.g. formula 3.322 in ref.[22], one gets

$$F(t, x) = \sqrt{\frac{\pi}{4t}} e^{-(2mt+x)^2/4t} \left[ 1 - \phi\left(\frac{2mt+x}{2\sqrt{t}}\right) \right], \tag{B.13}$$

where

$$\phi(x) = \frac{2}{\sqrt{\pi}} \int_0^x e^{-t^2} \mathrm{d}t \tag{B.14}$$

is the error function.

*Vanishing background field*

The one-dimensional kernels of the previous paragraph provide the building blocks for the 4 dimensional heat kernels, in the case of vanishing background field. We define

$$K_t(x, x' | \Delta_0)^{ab} \Big|_{B=0} = \delta^{ab} H_t(x, x'),$$

$$K_t(x, x' | \Delta_1)^{ab}_{\mu\nu} \Big|_{B=0} = \delta^{ab} H_t(x, x')_{\mu\nu}, \tag{B.15}$$

$$\mathcal{K}_t(x, x' | \Delta_2)_{si\alpha; tj\beta} \Big|_{B=0} = \delta_{st} \delta_{ij} \mathcal{K}_t(x, x')_{\alpha\beta},$$

with the r.h.s. further specified as follows

$$H_t(x, x')_{\mu\nu} = K_t^{T^3}(\mathbf{x}, \mathbf{x}') \left[ \delta_{\mu 0} \delta_{\nu 0} K_t^{\mathrm{NN}}(x_0, x_0') + \delta_{\mu k} \delta_{\nu k} K_t^{\mathrm{DD}}(x_0, x_0') \right],$$

$$H_t(x, x') = K_t^{T^3}(\mathbf{x}, \mathbf{x}') K_t^{\mathrm{DD}}(x_0, x_0'),$$

$$\mathcal{K}_t(x, x')_{\alpha\beta} = (-\slashed{\partial} + m)_{\alpha\alpha'} e^{-tm^2} K_t^{T^3}(\mathbf{x}, \mathbf{x}') \left[ (P_+)_{\alpha'\beta} K_t^{\tilde{\mathrm{N}}\mathrm{D}}(x_0, x_0') \right. \tag{B.16}$$

$$\left. + (P_-)_{\alpha'\beta} K_t^{\mathrm{D}\tilde{\mathrm{N}}}(x_0, x_0') \right].$$

# Appendix C

The asymptotic small $t$ expansion of the function $I(f, ts_1, ts_2)$ [eqs.(5.9),(5.11)] is obtained by expanding the heat kernels in powers of the background field.

*Expansion in powers of the background field*

The free quark propagator in the presence of the smooth background gauge field $B_\mu(x)$ satisfies

$$S(x, x') = S_0(x, x') - \int d^4x_1 \, S(x, x_1) \slashed{B}(x_1) S_0(x_1, x'), \tag{C.1}$$

where $S_0(x, x')$ denotes the free quark propagator for vanishing background gauge field. When this equation is iterated,

$$S(x, x') = S_0(x, x') + \sum_{n=1}^{\infty} S^{(n)}(x, x'), \tag{C.2}$$

the $n$-th order term is given by

$$S^{(n)}(x, x') = (-1)^n \int d^4x_1 d^4x_2 \cdots d^4x_n \, S_0(x, x_1) \\ \times \slashed{B}(x_1) S_0(x_1, x_2) \slashed{B}(x_2) \cdots \slashed{B}(x_n) S_0(x_n, x'). \tag{C.3}$$

This expansion of the propagator is easily turned into an expansion of the corresponding heat kernel. With the notation

$$S^{(n)}(x, x') = \int_0^\infty dt \, \mathcal{K}_t^{(n)}(x, x' | \Delta_2),$$
$$S_0(x, x') = \int_0^\infty dt \, \mathcal{K}_t(x, x'), \tag{C.4}$$

we obtain for the $n$-th order term

$$\mathcal{K}_t^{(n)}(x, x' | \Delta_2) = (-t)^n \int_0^1 ds_1 \cdots ds_{n+1} \, \delta\Big(\sum_{j=1}^{n+1} s_j - 1\Big) \int d^4x_1 d^4x_2 \cdots d^4x_n \\ \times \mathcal{K}_{ts_1}(x, x_1) \slashed{B}(x_1) \mathcal{K}_{ts_2}(x_1, x_2) \slashed{B}(x_2) \cdots \slashed{B}(x_n) \mathcal{K}_{ts_{n+1}}(x_n, x'). \tag{C.5}$$

Proceeding similarly for the gluon propagator, we obtain the corresponding expansion

$$I(f, t_1, t_2) = \sum_{m,n=0}^{\infty} I^{(m,n)}(f, t_1, t_2), \tag{C.6}$$

where the superscript $(m, n)$ stands for $m$ insertions in the gluon kernel and $n$ insertions in the quark kernel.

*Vanishing background field*

We start with the evaluation of the lowest order terms. Leaving the $\gamma$ algebra $D$-dimensional, contracting the color indices (appendix A), and using the notation of appendix B for the heat kernels, the integrand is given by

$$I^{(0,0)}(f, t_1, t_2) = -C_F \int d^4x\, d^4x'\, \text{tr}\,\{f(x,x') H_{t_1}(x,x')_{\hat{\mu}\hat{\nu}} \gamma_{\hat{\mu}} \mathcal{K}_{t_2}(x,x') \gamma_{\hat{\nu}}\}. \qquad (C.7)$$

As explained in appendix B, the kernel for the quark field is not known explicitly, even for vanishing background field. However, if the mass is set to zero, an explicit expression of the kernel can be obtained. Hence, for zero mass, the result to order $t^{-2}$ reads

$$\begin{aligned}
I^{(0,0)}(f, ts_1, ts_2) \stackrel{t \to 0}{\sim}\ & \frac{C_F}{16\pi^2 t^2 (s_1+s_2)^3} \Bigg[ s_1(1-\varepsilon) \int d^4x \\
& \times \text{tr}\,\{\not{\partial}^{(1)} f(x,x) - \not{\partial}^{(2)} f(x,x)\} \\
& + s_1(1-\varepsilon) \int d^3\mathbf{x}\, \text{tr}\,\{f(0,\mathbf{x},0,\mathbf{x}) + f(T,\mathbf{x},T,\mathbf{x})\} \\
& - 2(s_1+s_2)(1-\varepsilon/2) \int d^3\mathbf{x}\, \text{tr}\,\{f(0,\mathbf{x},0,\mathbf{x}) + f(T,\mathbf{x},T,\mathbf{x})\} \Bigg].
\end{aligned} \qquad (C.8)$$

To determine the influence of the mass we now assume the test function to have support only near the boundary at $x_0 = 0$. We may let $T$ become infinite without changing the asymptotic behavior near $x_0 = 0$. In this limit, the heat kernel representation of the quark propagator is again known (appendix B). It turns out that there is only 1 additional contribution at the order $t^{-2}$, so that the result for the massless case (C.8) is to be replaced by

$$\begin{aligned}
I^{(0,0)}(f, ts_1, ts_2) \stackrel{t \to 0}{\sim}\ & I^{(0,0)}(f, ts_1, ts_2)\big|_{m=0} \\
& + \frac{C_F}{16\pi^2 t^2} \frac{m(2\varepsilon - 4)}{(s_1+s_2)^2} \int d^4x\, \text{tr}\, f(x,x).
\end{aligned} \qquad (C.9)$$

*Effect of the background field*

In order to identify the potentially divergent diagrams, it is useful to follow Symanzik [1] and split the propagators and the corresponding heat kernels in a free space ($f$) and a surface ($s$) part. By definition, the free space propagator is the propagator which would be obtained if periodic boundary conditions had been imposed in all directions. It is singular if its arguments approach each other, while the surface propagator has a singularity if its arguments are close to each other and to one of the boundaries. One may set up a systematic power counting [23], and a moment of thought reveals the asymptotic behavior of $I^{(m,n)}(f, ts_1, ts_2)$,

$$I^{(m,n)}(f, ts_1, ts_2)^{(f)} \overset{t \to 0}{\sim} O(t^{-5/2+(m+n)/2}),$$
$$I^{(m,n)}(f, ts_1, ts_2)^{(s)} \overset{t \to 0}{\sim} O(t^{-2+(m+n)/2}). \quad (C.10)$$

Here, $(f)$ denotes the contribution with free space kernels only, while $(s)$ applies to diagrams with at least 1 surface kernel.

We see here that indeed very few diagrams may give contributions at order $t^{-2}$. Besides the case of vanishing background fields which has already been treated, there are only 2 further contributions involving the background field. Furthermore these are the same as in free space and thus cannot contribute to the boundary divergences.

It remains to calculate the 2 contributions for a single insertion in either the gluon or the quark kernel. Straightforward calculation leads to

$$I^{(1,0)}(f, ts_1, ts_2) \overset{t \to 0}{\sim} -\frac{N}{16\pi^2 t^2} \frac{s_1(1-\varepsilon)}{(s_1+s_2)^3} \int d^4x \, \text{tr}\{f(x,x)\slashed{B}(x)\},$$
$$I^{(0,1)}(f, ts_1, ts_2) \overset{t \to 0}{\sim} \frac{1}{N} \frac{1}{16\pi^2 t^2} \frac{s_2(1-\varepsilon)}{(s_1+s_2)^3} \int d^4x \, \text{tr}\{f(x,x)\slashed{B}(x)\}, \quad (C.11)$$

where terms of the order $t^{-3/2}$ have been neglected.

*Final result*

When the equations (C.8),(C.9) and (C.11) are combined, the asymptotic expansion for $I(f, t_1, t_2)$ itself is obtained. The result can be written in the form

$$I(f, ts_1, ts_2) \overset{t \to 0}{\sim} A(f, s_1, s_2) t^{-2} + O(t^{-3/2}), \quad (C.12)$$

where the coefficient function $A(f, s_1, s_2)$ is known explicitly and obtained through

$$A(f, s_1, s_2) = \lim_{t \to 0}\{t^2 \left[I^{(0,0)}(f, ts_1, ts_2) + I^{(1,0)}(f, ts_1, ts_2)\right. \\ \left. + I^{(0,1)}(f, ts_1, ts_2)\right]\}. \quad (C.13)$$